\documentstyle[epsf]{espart}
\begin{document}
\begin{frontmatter}
\title{Role of anharmonicities and non-linearities in 
heavy ion collisions. 
A microscopic approach.}

\author[Catania]{E.G. Lanza}, 
\author[Sevilla]{M.V. Andr\'{e}s}, 
\author[Catania]{F. Catara}, 
\author[Ganil]  {Ph. Chomaz} and 
\author[Ganil]  {C. Volpe}
\address[Catania]{ Dipartimento di Fisica Universit\'a di Catania and 
  INFN, Sezione di Catania, I-95129 Catania, Italy}
\address[Sevilla]{ Departamento de F\'{\i}sica At\'omica, Molecular y
  Nuclear, Universidad de Sevilla, Apdo 1065, 41080 Sevilla, Spain}
\address[Ganil]{ GANIL, B.P. 5027, F-14021 Caen Cedex, France}

\begin{abstract}
Using a microscopic approach beyond RPA  to treat anharmonicities, we
mix two-phonon states among themselves and with one-phonon states. We
also introduce non-linear terms in the external field. These non-linear
terms and the anharmonicities are not taken into account in the
"standard" multiphonon picture.  Within this framework we calculate
Coulomb excitation of $^{208}$Pb and $^{40}$Ca by a  $^{208}$Pb  nucleus
at 641 and 1000MeV/A. We show with different examples the importance of
the non-linearities and anharmonicities for the excitation cross
section. We find an increase of 10\% for $^{208}$Pb and 20\% for
$^{40}$Ca of the excitation cross section corresponding to the energy
region of the double giant dipole resonance with respect to the 
"standard" calculation. We also find important effects in the low energy
region. The predicted cross section in the DGDR region is found to be
rather close to the experimental observation.
\end{abstract}
\end{frontmatter}

\section{Introduction}           

States that can be interpreted as the first quanta of collective
vibrations  are a general property of quantum mesoscopic systems which
can be found in  various fields of physics. In nuclear physics, such
vibrational states of the nucleus  have been known for many
years~\cite{bm}. These one-phonon states are  present both in the
low-lying excitation spectra of nuclei and at higher energies. The
latter are the Giant Resonances (GR). The existence of  two-phonon
states, i.e. states which can be described as double excitations of 
elementary modes, has also been predicted since the early days of the 
collective model~\cite{bm}. Such states have been observed long time ago
in the low-lying spectra. More recently, two-phonon states built with
giant resonances have been populated in heavy ion inelastic
scattering~\cite{f88}, in double charge exchange ($\pi^{\pm},\pi^{\mp}$)
reactions~\cite{mor88} and in Coulomb excitation at high
energy~\cite{schm,ri93,ST95}.  For a review, see  ref.~\cite{cho95}. 

In the harmonic approximation, these states are predicted as degenerate
multiplets  located at an excitation energy equal to the sum of the
individual phonon energies.  When the residual interaction is taken into
account, the degeneracy is broken by the coupling  between phonons. In
the present article we consider the residual interaction  of two-phonon
states among themselves and with one-phonon states. Therefore, the
eigenstates are linear combinations  of one- and two-phonon components
while the energies are shifted and splitted with respect to the harmonic
limit. We will call such states mixed states. Evidence of such
anharmonic behaviour can be found, for example, in a
$(\gamma,\gamma\prime)$ experiment ~\cite{plb94} where the observation
of large dipole strength in the low-lying spectrum of some ${\rm Sn}$
isotopes is reported.  Such strength is interpreted by the authors as
due partially to the population of the $1^-$ member of the quintuplet of
states based on the  $| 2^+\otimes 3^->$ two-phonon state, and partially
to the admixture of the (one-phonon) GDR in the wavefunction of the
state observed around 3.5 MeV excitation energy. As it has been shown
in~\cite{cat89}, the inclusion of the residual interaction among
two-phonon states leads to small, but sizeable, anharmonicities also in
the high-lying spectrum, namely for those states that in the harmonic
limit are described as double excitations of GR. The mixing between one-
and  two-phonon states further increases the anharmonicities. When an
external field acts on a nucleus, it excites the eigenstates of the
internal  hamiltonian, which, in our approach,  are superpositions of
one- and two-phonon states. 

The microscopic theory suited for the description of collective
vibrational  states is the Random Phase Approximation (RPA). Two-phonon
states and their  mixing among themselves and with one-phonon states can
be generated by using boson mapping techniques and by taking into
account terms of the residual interaction which do not enter at the RPA
level~\cite{cat89,bea92}. In this way one has an RPA based approach to
treat anharmonicities.

In a nucleus-nucleus collision, the mutual excitation of the two
partners is described as due to the action of the mean field of each
nucleus on the other one, i.e. by a one body operator. Assuming that it
induces small deformations of the density, only the particle-hole (ph)
terms of the  external  mean field are usually taken into account. This
amounts to consider as elementary processes only those corresponding to
the creation or annihilation of one phonon. In this  approximation, the
external field is linear in the creation and annihilation operators of
phonons. When the particle-particle (pp) and hole-hole (hh) terms of the
external field are also included, the direct excitation from the ground
state to two-phonon states as well as the transition between  one-phonon
states become possible. These terms can be expressed as quadratic in the
creation and  annihilation operators of phonons   and so correspond to
non-linear terms in the excitation operator.

In the "standard" approach, based on the independent multiphonon
picture, the effects coming from both anharmonicities and
non-linearities are neglected (see for instance ref.\cite{be96}). 
Recent experimental data on Coulomb excitation at relativistic  energies
have raised some questions on the adequacy of that picture. Indeed, in
the excitation of $^{136}$Xe on $^{208}$Pb, the experimental cross
section to the double GDR (DGDR) has been found to be 2 to 4 times
larger than the theoretical one~\cite{schm}. Recently, new experimental
results ~\cite{ST95} on the excitation of several nuclei have shown that
the disagreement ranges from about 10\% to 60\%, being about 30\% in the
case of $^{208}$Pb.  In a previous paper~\cite{vol95},  by using a 
one-dimensional  oscillator model to mimick nuclear states, we have
shown that the effects of anharmonicities and non-linearities can lead
to an important enhancement of the cross section in the energy range 
around twice that of the GDR. In this model neither spin nor parity
were taken into account. Besides, only one type of phonons was
considered. In the present paper we present more realistic 
calculations, where the collective states of the target nucleus and the
action on it of the Coulomb field of the projectile are described
starting from RPA. Anharmonicities and non-linearities are included by
means of boson mapping techniques. Both low-lying collective states and
giant resonances are considered as elementary phonons.

We have done calculations for the $^{208}$Pb$ + ^{208}$Pb system at 641
and 1000 MeV per nucleon for which experimental data exist ~\cite{ST95}. 
We have also studied the Coulomb excitation of $^{40}$Ca in the reaction 
$^{208}$Pb$ + ^{40}$Ca at 1000 MeV/A although there are no experimental 
data for this case.  In both cases we consider as elementary modes all
natural-parity RPA phonons whose multipolarity is lower than 4 and whose 
contribution to the associated energy weighted sum rule (EWSR) is larger
than 5\%. Then, we have built the residual interaction in the one and
two-phonon space and we have diagonalized the hamiltonian in this
subspace in order to define the mixed states $|\phi_\alpha>$ .  By
solving the time dependent Schr\"odinger equation in this subspace  we
get the probability amplitudes for each of the $|\phi_\alpha>$ states
from which we calculate the cross section. We will describe in detail
the results for the $^{208}$Pb$ + ^{208}$Pb system at 641 MeV/A, the
results at 1000 Mev/A being essentially the same except for the absolute
values of the cross section which are higher in the latter case. We will
mostly discuss the two regions around the energy of the states built
with two low-lying states or with two giant resonances. We will see that
non-linearities and anharmonicities may strongly change the cross
sections associated with some specific states. Their influence
is found to be of the order of 10\% for $^{208}$Pb and 20\% for 
$^{40}$Ca in the DGDR energy region, bringing the theoretical results 
closer to the experimental ones.

In the next section we detail the model employed and in section 3
we describe the semiclassical electromagnetic field used to excite the
nuclei.  Section 4 is devoted to the description of the results on
$^{208}$Pb where we discuss in a detailed way the effects of both
anharmonicities and non-linear terms on the excitation cross
section. The results for $^{40}$Ca are reported in section 5 and finally
we draw our conclusions in section 6.

\section{The multiphonon picture}

Heavy ion collisions at high incident energies can be described within a
semiclassical approach, where the relative motion is treated classically
while quantum mechanics is used for the internal degrees of freedom of
the colliding nuclei. For grazing  and large impact parameter 
collisions the densities of the two nuclei have a small overlap.
Therefore, the total hamiltonian can be written as
\begin{equation}
H = H_A + H_B
\end{equation}
where $H_A (H_B)$ denotes the hamiltonian of nucleus A(B) and
\begin{equation}
H_A = H_A^0 + \sum_{\alpha \alpha^\prime} 
  <\alpha|U_B ({\bf R}(t))|\alpha^\prime> 
   a^\dagger_\alpha a_{\alpha^\prime}
  = H_A^0 + W_A (t)
\end{equation}
$H^0_A$ being the internal hamiltonian of A.  $W_A$ describes the
excitation of A by the mean field $U_B$ of nucleus B, whose matrix
elements  depend on time through the relative coordinate {\bf R}(t). The
sums over the single particle  states, denoted by $\alpha$ and 
$\alpha^\prime$, run over both particle and hole states.   

\subsection{Harmonic approximation}

Within RPA, the excited states $|\Psi_\nu>$ of each nucleus are 
described as superpositions of $p h$ and $h p$ configurations with
respect to  the ground state $|\Psi_0>$
\begin{equation}
|\Psi_\nu> = q^\dagger_\nu |\Psi_0>=\sum_{ph } 
    [X^\nu_{ph} a^\dagger_p a_h -
                    Y^\nu_{ph} a^\dagger_h a_p] |\Psi_0>
\end{equation}
where the amplitudes $X$ and $Y$ are solutions of the RPA secular
equation, with eigenvalues $E_\nu$. The ground state is defined as the
vacuum of the $q_\nu$ operators
\begin{equation}
q_\nu |\Psi_0>=0
\end{equation}
In order to avoid unnecessarily complicated expressions, we do not
introduce explicitly the coupling to total angular momentum and isotopic
spin. When the RPA phonons are mapped onto bosons~\cite{rs}, the
internal hamiltonian of the nucleus can be written as
\begin{equation}
H^0 = \sum_{\nu} E_\nu Q^\dagger_\nu Q_\nu
\end{equation}
which shows that the excitation spectrum is harmonic. The boson
operators $Q^\dagger_\nu$ and $Q_\nu$ in the above equation are given by
\begin{equation}
Q^\dagger_\nu =\sum_{ph } [X^\nu_{ph} B^\dagger_{ph} - Y^\nu_{ph} B_{ph}]
\end{equation}
with the same X and Y amplitudes as in eq.(3) but in terms of the boson
images ($B^\dagger_{ph}$ and $B_{ph}$) of the ph operators
\begin{equation}
a^\dagger_p a_h \rightarrow B^\dagger_{ph} + .......
\end{equation}
In the above equation we have indicated only the first term of the boson 
mapping.  Assuming that the  external field induces only small
deformations of the density, only $p h$ and  $h p$ terms contribute to
$W_A$ and one gets
\begin{equation}
W_A(t) = \sum_{ph} <p|U_B({\bf R}(t))|h> a^\dagger_p a_h + h.c.
\end{equation}
By introducing the boson mapping of eq.(7), it can be rewritten as
\begin{equation}
W_A(t) = \sum_{\nu} W_\nu^{10}(t) Q^\dagger_\nu + h. c.
\end{equation}
with
\begin{equation}
W_\nu^{10}  = <\Psi_\nu| W(t) |\Psi_0> 
\end{equation}

The independent multiphonon picture is based on eqs. (5) and (9). Within 
this picture, the  Schr\"odinger equation can be solved exactly and the
state of each nucleus at time t is found to be the coherent state
\begin{equation}
|\Phi(t)> = \prod_{\nu} e^{-{1 \over 2} |I_\nu (t)|^2}  \sum_{n_\nu}
  {[I_\nu (t)]^{n_\nu} \over n_\nu !} e^{-i n_\nu E_\nu t} 
   (Q^\dagger_\nu)^{n_\nu} |0>
\end{equation}
with
\begin{equation}
I_\nu = \int_{-\infty}^t W_\nu^{10} (t^\prime) 
e^{-i n_\nu E_\nu t^\prime} \, dt^\prime
\end{equation}                 
where the integral is performed along the relative motion trajectory 
corresponding to a definite impact parameter. The probability amplitude
to excite one- or two-phonon states is calculated by projecting eq. (11)
on the corresponding  states
\begin{equation}
|\nu> = Q^\dagger_\nu |0>
\end{equation}
\begin{equation}
|\nu\nu'> = (1 + \delta_{\nu\nu'})^{-1/2} Q^\dagger_\nu 
Q^\dagger_{\nu'} |0>
\end{equation}
where $|0>$ is the vacuum of the $Q^\dagger_\nu$ operators. Finally, the
cross sections are obtained by integrating the relevant  probabilities
over the impact parameter.

\subsection{Non-linear excitation}

In this section we present an approach to go beyond the independent 
multiphonon picture by eliminating its two main limitations, the first
in the external field and the second in  the internal hamiltonian. Let
us first consider the $p p$ and $h h$ contributions to the sums in
eq.(2), which are neglected in the linear approximation for the external
field. Assuming the same boson mapping, truncated at the lowest order,
it is easily shown~\cite{cat89} that the mappings
\begin{equation}
a^\dagger_p a_{p^\prime} \rightarrow 
     \sum_h B^\dagger_{ph}  B_{p^\prime h} 
\end{equation}
\begin{equation}
a_h a^\dagger_{h^\prime} \rightarrow 
     \sum_p B^\dagger_{ph}  B_{p h^\prime } 
\end{equation}
are exact, in the sense that they preserve the commutation relations
between fermion-pair operators.  Using these relations, the inclusion of
the $p p$ and $h h$  terms in  eq.(2) gives a W quadratic in the boson
operators  $B^\dagger_{p h}$ and $B_{p h}$.  By expressing the latter in
terms of the collective bosons  $Q^\dagger_\nu$ and $Q_\nu$ one gets
\begin{equation}
W = W^{00} + \sum_\nu W^{10}_\nu Q^\dagger_\nu +  h.c. +
  \sum_{\nu\nu^\prime} W^{11}_{\nu\nu^\prime} 
  Q^\dagger_\nu Q_{\nu^\prime} +
  \sum_{\nu\nu^\prime} W^{20}_{\nu\nu^\prime} Q^\dagger_\nu 
  Q^\dagger_{\nu^\prime} + h.c.  
\end{equation}
where
\begin{equation} \label{e:W10}
W^{10}_\nu = \sum_{ph} (W_{ph} X_{ph}^{\nu^*} + W_{hp} 
Y_{ph}^{\nu^*})
\ \ 
\end{equation}
is the standard linear response expression, whereas 
\begin{equation} \label{e:W11}
W^{11}_{\nu \nu'} = \sum_{php'h'} 
\left( W_{pp'} \delta_{hh'} - W_{hh'} \delta_{pp'} \right)
\left(  X_{ph}^{\nu^*} X_{p'h'}^{\nu'}+ Y_{ph}^{\nu^*}Y_{p'h'}^{\nu'} 
\right)
\ \ 
\end{equation}
\begin{equation} \label{e:W20}
W^{20}_{\nu \nu'} = \sum_{php'h'} 
\left( W_{pp'} \delta_{hh'} - W_{h'h} \delta_{pp'} \right)
 X_{ph}^{\nu^*} Y_{p'h'}^{\nu'^*}
\ \                            
\end{equation}
provide new excitation routes.

The matrix elements of $U_B$ depend on the considered excitation
mechanism.  Since the general discussion we present here is independent
of their form, we  postpone to the next section the derivation of their
expressions in the case of Coulomb excitation at relativistic energy.
                                           
The hamiltonian $H_A$, with the inclusion of the terms $W^{11}$ and
$W^{20}$, is a quadratic form in the $Q^\dagger_\nu$ and $Q_\nu$
operators. Therefore, a coherent state solution to the Schr\"odin\-ger
equation still exists. We do not exploit this property because it does
not hold any more when the anharmonicities are included, as we are going
to do in next subsection.  The effects of introducing non-linear terms
in the external field  can be important, for example, whenever some
selection rule disfavours one of the two steps necessary to make the
transition from the ground state to a two-phonon state through the
action of $W^{10}$ alone.  The term $W^{11}_{\nu \nu^\prime}$ describes
the transition from the one-phonon state $|\Psi_\nu>$ to
$|\Psi_{\nu^\prime}>$ or from a two-phonon state to  another one. In
ref.~\cite{ca86,ca87,ca88} it was shown that these non-linear terms can
lead to an increase of the population of two-phonon states.  The term
$W^{20}$ induces a direct transition from the ground state to a 
two-phonon state that can be very important. This effect has been
already  reported  in~\cite{plb94} where the direct matrix element
between the ground state and the dipole member of the low-lying $|
2^+\otimes 3^->$ quintuplet of states in some Sn isotopes was found to
be very large. Similar results, but involving double GR states, have
been obtained in~\cite{ccg92}.

\subsection{Anharmonic spectrum}

Let us now turn our attention to the other limitation of the independent
multiphonon picture we have stressed above, namely the assumption that
the internal hamiltonian has the harmonic form of eq.(5). The simplest
way to go beyond this approximation starts from the observation that in 
RPA only the  $phph$ and $pphh$ terms of the residual interaction are
taken into account. The $pppp$ and $hhhh$ terms,  when expressed by the
same boson mapping used before, introduce a coupling between two-phonon
states~\cite{cat89} while the remaining, $ppph$ and $hhhp$, terms mix
one- and two-phonon states.  Finally, when considering two-phonon states
one should also take  care of the possible violation of the Pauli
principle. In the boson mapping method the exclusion principle is
introduced through high order terms in the boson expansion ~\cite{cat89}
built to conserve the  fermion-pair commutation algebra.  In such a way
an additional residual interaction between two-phonon states  coming
from the particle-hole matrix elements is generated \cite{bea92}. As a
result of these different couplings, the eigenstates of the internal
hamiltonian of each nucleus are
\begin{equation}
|\Phi_\alpha> = \sum_\nu c^\alpha_\nu |\nu> + \sum_{\nu_1 \nu_2}
      d^\alpha_{\nu_1 \nu_2} |\nu_1 \nu_2 >
\end{equation}
Therefore, the states excited by the external field will be such mixed
states and one cannot speak of pure one- or two-phonon excitations any
more.  However, when a $|\Phi_\alpha>$ state has a strong overlap  with
a one-phonon state we may discuss the associated cross-section as part
of the one-phonon cross-section. Conversely, if the $|\Phi_\alpha>$
state is dominated by its two-phonon components we may speak about
two-phonon  strength.  

For example, let us consider a state  $|\Phi_\alpha>$  which strongly
overlaps with a two-phonon state, i.e. whose largest component is
$|\nu_1 \nu_2>$.  In addition to the possible excitation of
$|\Phi_\alpha>$  via this two-phonon component,  this state can also be
excited by  $W^{10}$ through its one-phonon components. However, the 
energy  $E_\alpha$ of  $|\Phi_\alpha>$ will be not far from $E_{\nu_1} +
E_{\nu_2}$. Therefore, it will contribute to the cross section at that
energy. In this sense, because of its structure and of its energy,  one 
may say that it contributes to the two-phonon cross section.  This fact
was disregarded in  ref.~\cite{bro}, where the mixing of a huge number
of one- and two- phonon states was considered. A good description of the
width of the GDR was thus achieved. However,  all the $|\Phi_\alpha>$
states were considered to be one-phonon states when they were excited
through their one-phonon components while the two-phonon excitations
were calculated  in~\cite{bro} as the transition to states of the form 
$|\Phi_\alpha \otimes \Phi_{\alpha^\prime}>$.  Therefore, that
calculation is somewhat equivalent to consider a harmonic spectrum  with
the states  $|\Phi_\alpha>$ as the elementary quanta. 

\subsection{Time-dependent excitation process}

The cross section is calculated, non perturbatively, by  solving the
Schr\"odinger equation in the space of the ground state and the
$|\Phi_\alpha>$ states. Then the time dependent state, $|\Psi(t)>$,  of
the nucleus can be expressed as 
\begin{equation}
|\Psi(t)>  = \sum_\alpha A_\alpha (t) e^{-i E_\alpha t} |\Phi_\alpha>
\end{equation}
where the ground state is also included in the sum as the term 
$\alpha=0$. The amplitudes  $A_\alpha (t)$ are solutions of the set of
linear differential equations
\begin{equation} \label{adot}
\dot A_\alpha (t) = -i \sum_{\alpha \alpha^\prime} e^{i (E_\alpha
    - E_{\alpha^\prime}) t} <\Phi_\alpha|W(t)|\Phi_{\alpha^\prime}> 
\end{equation}
and the probability of exciting the internal state  $|\Phi_\alpha>$ is
given by
\begin{equation}
P_\alpha = |A_\alpha (t = + \infty)|^2
\end{equation}
for each impact parameter. Finally, by integrating $P_\alpha$ over the
impact parameter we obtain the cross section 
\begin{equation}
\sigma_\alpha = 2\pi \int_{0}^{+\infty} P_\alpha (b) T(b) b db,
\end{equation}
where the transmission coefficient $T(b)$ has been taken equal  to a
sharp cutoff function $\theta (b-b_{min})$. The parameter $b_{min}$ is
usually chosen such that the contribution from the nuclear part can be
neglected.

\section{ Relativistic Coulomb excitation}

Let us now look in detail the multipole expansion  of the external field
of eq.(2). Alder and Winther \cite{AW79} have worked out an analytic
expression for the Fourier transform of the semiclassical 
electromagnetic field  in relativistic nucleus-nucleus collisions, with
the assumptions that the projectile follows a straight-line trajectory
and that the charge densities of both nuclei do not overlap.  Therefore,
to get the time dependence of the  electromagnetic coupling potential
the inverse  Fourier transform of the expressions derived in \cite{AW79}
can be taken.  This procedure has the advantage that the multipole
expansion of the time dependent coupling potential is readily known as
well as its electric and magnetic components.

Let us introduce the Fourier components of the  time dependent coupling
potential 
\begin{equation} \label{e1}
W(t) = {1 \over {2\pi}} 
\int_{-\infty}^{+\infty} e^{-i\omega t} W(\omega) d\omega 
\end{equation}
\noindent Introducing the expansion of the external field in multipoles 
$W^{\lambda\mu}$  
\begin{equation} \label{e2}
W(t) 
={1\over {2\pi}}
\sum_{\lambda,\mu} \int_0^{+\infty} \big( e^{i\omega t}
(-1)^{\lambda +\mu}+ e^{-i\omega t} \big) 
W^{\lambda\mu}(\omega ) d\omega
\end{equation}
where we have taken into account the behaviour of the  multipoles
$W^{\lambda\mu}$ for negative $\omega $.

It is shown in ref. \cite{AW79} that  the contribution to $W(\omega)$ of
the $(\lambda,\mu )$ multipole can be expressed in terms of electric
($\pi=E$) and magnetic ($\pi=M$)  one-body operators 
\begin{equation} \label{a1}
W^{\lambda \mu} (\vert \omega \vert) = 
{{Z_p e} \over { v\gamma}}
\sum_\pi G_{\pi \lambda \mu}\big({c\over v}\big) (-1)^\mu K_\mu 
(\beta \omega ))
{{\sqrt{2\lambda+1}}} \big({\omega \over c}\big)^\lambda 
{\cal M}_t(\pi\omega \lambda -\mu)
\end{equation}
\noindent where $\beta\omega$  is the adiabaticity parameter related to
the impact parameter $b$ and to the Lorentz contraction factor,
$\gamma$,  and where  $K_\mu$ are the modified Bessel functions. The
expressions of the functions $G_{\pi \lambda \mu}$ can be found in ref.
\cite{AW79}.

The 2$^\lambda $-pole electric transition operator is given by
\begin{equation} \label{m1}
{\cal M} (E\omega \lambda \mu)= 
{{(2\lambda+1)!! c^\lambda} \over {\omega^{\lambda+1} (\lambda+1)}}
\int {\bf J}({\bf r})\cdot {\bf \nabla}\wedge {\bf L}
(j_\lambda({{\omega r} \over c})Y_{\lambda\mu}(\hat{r})) d^3{ r}
\end{equation}
\noindent where ${\bf J}$ is the current density operator while
$j_\lambda$ is a spherical Bessel function. This operator can also be
written down \cite{rs} as
\begin{eqnarray} \label{m2}
{\cal M} &(&E\omega \lambda \mu)= 
{{(2\lambda+1)!! c^\lambda} \over {\omega^\lambda (\lambda+1)}}
\nonumber \\
&\times&\int \left\{ {\bf \rho} ({\bf r}) Y_{\lambda\mu}(\hat{r}) 
{\partial \over {\partial r}} (rj_\lambda({{\omega r} \over c})) + 
i {\omega \over c^2} {\bf J}({\bf r})\cdot {\bf r}
Y_{\lambda\mu}(\hat{r})j_\lambda({{\omega r} \over c}) \right\} d^3{ r}
\end{eqnarray}
\noindent where the charge density operator ${\bf \rho}({\bf r}) $ has
been introduced. The second term will be neglected since, relative to
the first, it is of the  order of $\hbar\omega /2 m_p c^2$. 

To get the time dependent coupling potential as  the Fourier transform
(\ref{e2}-\ref{a1}),  we need the transition operators at any value of
$\omega$. Since the argument of the spherical Bessel function in the
operator is $\omega r/c$, the  dependence on $\omega$ and r of the
multipole $W^{\lambda \mu}$  will not factorize. Therefore there would
be no factorization of the time and r dependence in the coupling
potential. However, in the limit of long wavelengths, the first term in
expression (\ref{m2}) reduces to  the well-known static electric
multipole operator,
\begin{equation} \label{m3}
{\cal M} (E\omega \lambda \mu) \simeq \hat{Q}_{\lambda\mu} =
\int {\bf \rho}({\bf r})  r^\lambda Y_{\lambda\mu}(\hat{r}) d{\bf r}
\end{equation}
\noindent which does not depend on $\omega$. In a similar way, the
general magnetic operator
\begin{equation} \label{mm1}
{\cal M} (M\omega \lambda \mu)= 
-i{{(2\lambda+1)!! c^{\lambda-1}} \over {\omega^{\lambda} (\lambda+1)}}
\int {\bf J}({\bf r})\cdot {\bf L}
(j_\lambda({{\omega r} \over c})Y_{\lambda\mu}(\hat{r})) d{\bf r}
\end{equation}
\noindent in the limit of long wavelengths, becomes
\begin{equation} \label{mm3}
{\cal M} (M\omega \lambda \mu) \simeq \hat{M}_{\lambda\mu} =
{1 \over {c(\lambda +1)}}
\int ({\bf r} \wedge {\bf J}({\bf r}))\cdot ({\bf \nabla}
r^\lambda Y_{\lambda\mu}(\hat{r})) d{\bf r}
\end{equation}
That means that, in the limit of long wavelengths, neither  the electric
nor the magnetic transition operators depend on $\omega$ (we will
therefore omit $\omega$ in the arguments of ${\cal M}$) and  they will
come out of the integral in~(\ref{e2}). We just need to know
\begin{equation} 
H_{\lambda \mu}(\beta,t)=\int_0^{+\infty} \big( e^{i\omega t}
(-1)^{\lambda +\mu}
+ e^{-i\omega t} \big) \omega^\lambda K_\mu (\beta\omega) d\omega
\end{equation}
\noindent with $\mu \geq 0$. $H_{\lambda \mu}(\beta,t)$ is an analytic
function whose explicit expression can be found in the appendix.

Therefore, in the long wavelength limit, the following analytic
expression of  the time dependent coupling potential
\begin{eqnarray}\label{multi}
W(t) 
&=&{{Z_p e} \over {2\pi v\gamma}}
\sum_{\pi \lambda \mu} G_{\pi \lambda \mu}\big({c\over v}\big) (-1)^\mu 
{{\scriptstyle{\sqrt{2\lambda+1}}} \over {c^\lambda}} 
H_{\lambda \mu}(\beta,t)
\ {\cal M}(\pi \lambda -\mu) \nonumber \\
&=&\sum_{\pi \lambda \mu} g_{\pi \lambda \mu}(\beta,t)
(-1)^\mu {\cal M}(\pi \lambda -\mu)/e
\end{eqnarray}
\noindent can be explicitly derived.  From the use of the static
multipole operators has followed that any of the term in the sum
factorize into two elements, the first one depending on collisions
properties, the second one acting  on the nucleus being excited.

If one considers only the electric components, which are the most
important, only natural parity states can be excited in a first order
calculation. However, in a coupled channel calculation, as the present
one, non natural parity states have to be included since they can be
reached, for example, through a two step process. However we will see in
the following that these contributions remain small.

We want to have a feeling about the limits that the use of the static
electric  operator imposes on the interpretation of our results. If a
first-order harmonic and linear calculation was to be done, we could
just compare  the complete matrix elements $<I_f\vert \vert{\cal M}(E
\omega_{if}\lambda)\vert\vert I_i>$ and  the approximate one  $<I_f\vert
\vert \hat{Q}_\lambda \vert \vert I_i>$.  The matrix elements connecting
the ground state with one phonon states are consistent within a maximum
of a  few per thousand for low-lying states, and differ by a few percent
when the GDR or the ISGQR are considered.

\begin{figure} [h]
\begin{center}
\mbox{{\epsfxsize=13truecm \epsfysize=9truecm \epsfbox{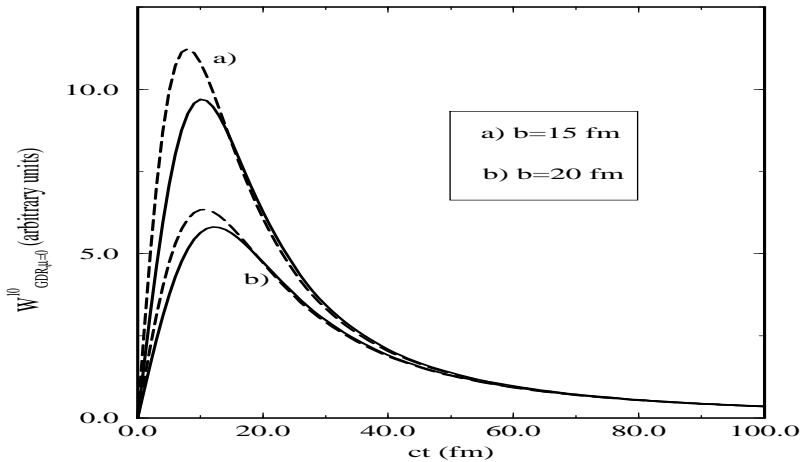}}}
\end{center}
\caption 
{Matrix element of the coupling potential between the ground state of
$^{208}$Pb  and its GDR with  magnetic quantum number zero, as function
of time.  There are two groups of lines corresponding to two different
impact parameters. The solid lines have been obtained using the general
expression (see eq. 30) of the electric dipole operator, while to get
the dashed line the static expression (eq. 31)  has been used.}
\label{fg1}
\end{figure}

\begin{figure} 
\begin{center}
\mbox{{\epsfxsize=13truecm \epsfysize=9truecm \epsfbox{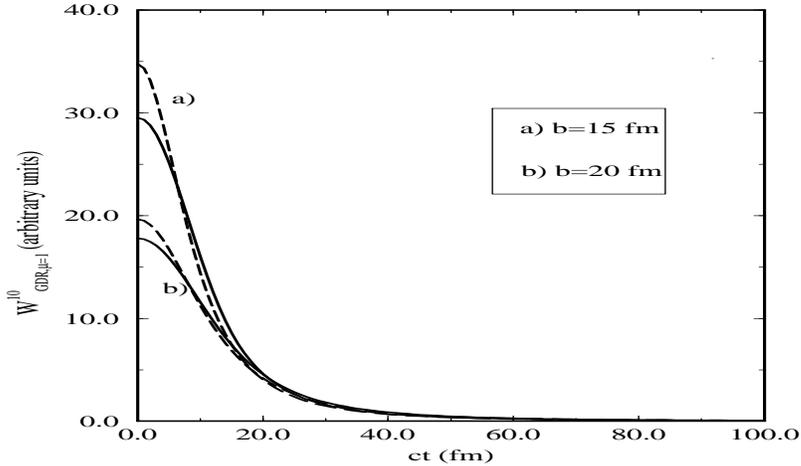}}}
\end{center}

\caption 
{As figure \ref{fg1}, but for magnetic quantum number one.}
\label{fg2}
\end{figure}

\begin{figure} 
\begin{center}
\mbox{{\epsfxsize=13truecm \epsfysize=9truecm \epsfbox{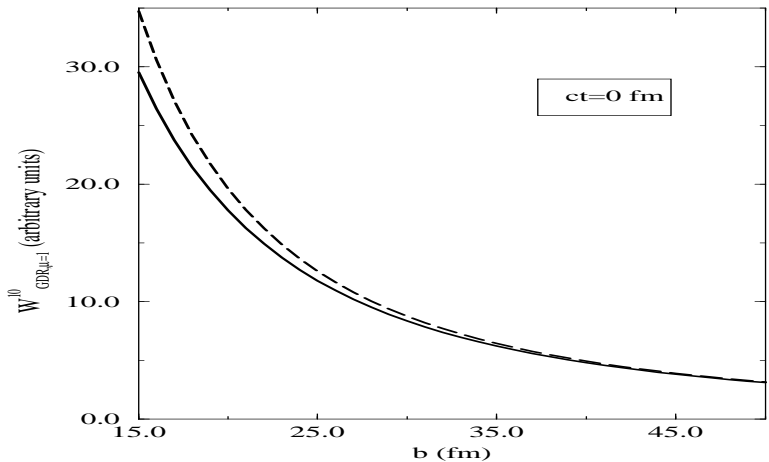}}}
\end{center}

\caption 
{Impact parameter dependence of the matrix element of the coupling
potential  at time t=0 between the ground state of $^{208}$Pb and its
GDR with magnetic quantum number one.}
\label{fg3}
\end{figure}

In a coupled-channel calculation not just a fixed $\omega_{if}$, but the
full range of $\omega$ values will contribute to the Fourier transform
(\ref{e1}- \ref{a1}). 
As an illustration, let us consider the colliding system 
$^{208}$Pb+$^{208}$Pb at E$_{lab}$= 641 MeV 
to compare the  exact multipole expansion with
the long wavelength limit. The associated time-dependent transition
matrices from the ground state to the giant dipole resonance, 
$<GDR,\mu\vert W(t) \vert 0>$, are presented in figure \ref{fg1} for
the magnetic quantum number $\mu=0$ and in figure \ref{fg2} for $\mu=1$
for two different values of the impact parameter. The solid lines
correspond to calculations in which the general expression of the
electric multipole operator has been taken into account, and the inverse
Fourier transform of the corresponding amplitude has been carried on
numerically. The dashed lines correspond to the use of the static
electric operator and the analytic expression (\ref{multi}). We can see
that qualitatively the time dependence is well reproduced, while the
quantitative agreement gets better as the impact parameters increases.
That is essentially due to the adiabatic cutoff that the modified Bessel
function $K_\mu(\beta \omega)$ introduces. This function decays
exponentially when the argument becomes bigger than 2
~\cite{Ab}. Therefore, if the impact parameter increases the relevant
range of $\omega$ in the integral is reduced and we get closer to the
long wavelength limit expression for the interacting potential. This
effect can be seen in figure \ref{fg3} where the behaviour with the
impact parameter of $<GDR,\mu=1\vert W(t=0) \vert 0>$  is presented
at t=0. This is the time at which the difference between both approaches
is maximum when $\lambda+\mu$ is even. A similar behaviour is found when
the matrix elements $W^{11}$ or $W^{20}$ are considered.

The  conclusion of this study is that,  in the excitation energy region
we will consider, it is reasonable to use the static electric operator.
This amounts to a considerable saving of calculations since, for each
multipole, the  time and the r dependence factorize.

\section{Results about the excitation of $^{208}$Pb}

Let us now apply the above formalism to a specific nucleus, namely the 
$^{208}$Pb excited in a collision with a Pb nucleus at 641 MeV per
nucleon. We will first discuss the effect of the anharmonicities on the
RPA spectrum. Secondly, we will look at the effect of non-linear terms
in the external field. Finally, we will consider the influence of both
these terms on the excitation probability and the cross sections. 

\subsection{Energy spectrum}

The one-phonon basis is calculated in the self-consistent RPA with SGII
Skyrme interaction \cite{SGII}. Although we are using an explicit
neutron  proton representation the isospin results to be a rather good
quantum number as far as collective states are concerned. We have
selected all the states which exhaust at least $5\%$ of the  appropriate
EWSR and, for a particular spin and parity (and isospin), we have
grouped together the ones which are closer in energy according to the
method described in ref.~\cite{cat89}.  We have considered the
one-phonon states reported in table \ref{fpb},  
i.e. the various components of
the isoscalar monopole resonance (GMR),  the components of the isovector
dipole resonance  $(GDR)$, the low-lying $2^+$ state and the quadrupole
resonances, both isoscalar (ISGQR) and isovector (IVGQR), and finally
the collective low-lying $(3^-)$ and high-lying $(HEOR)$ isoscalar 
octupole states. 

\begin {table} 
\caption { One-phonon basis for the nucleus $^{208}$Pb. For each state
its spin and parity, isospin, energy and percentage of the EWSR are
reported.}
\label{fpb}

\begin{tabular}{||r||cc|r|r||} \hline
 Phonons  &$J^\pi$&$   T   $&$ E (MeV) $&$ \% EWSR$\\ \hline \hline
$ GMR_1  $&$ 0^+ $&$   0   $&$ 13.610  $&$ 61 $\\
$ GMR_2  $&$ 0^+ $&$   0   $&$ 15.022  $&$ 28 $\\ \hline
$ GDR_1  $&$ 1^- $&$   1   $&$ 12.435  $&$ 63 $\\ 
$ GDR_2  $&$ 1^- $&$   1   $&$ 16.662  $&$ 17 $\\ \hline
$ 2^+    $&$ 2^+ $&$   0   $&$  5.545  $&$ 15 $\\
$ ISGQR  $&$ 2^+ $&$   0   $&$ 11.599  $&$ 76 $\\ 
$ IVGQR  $&$ 2^+ $&$   1   $&$ 21.815  $&$ 45 $\\ \hline
$ 3^-    $&$ 3^- $&$   0   $&$  3.464  $&$ 21 $\\
$ HEOR   $&$ 3^- $&$   0   $&$ 21.302  $&$ 37 $\\ \hline
\end{tabular}

\end{table}

We have then constructed the residual interaction between the one- and
two-phonon states and also among the two-phonon states. The two-phonon
states are coupled to a total angular momentum and parity. In the case
of the $1^-$ states,  while the coupling between one- and two-phonon
states is of the order  of 1/2 MeV up to 1 MeV, the coupling between
two-phonon states is, in average, about one order of magnitude smaller.

\begin {table} 
\caption { Characteristics of the $|\phi_\alpha>$ dipole 1$^-$ states 
resulting from the diagonalization of the internal hamiltonian. In the
first column  we indicate the dominant component. The values in the
second column remind us the energies associated with this component in
the harmonic approach. The shift in the energy produced by the
anharmonicities is indicated by $\Delta E$ (in KeV).  We can compare
these values with the diagonal matrix elements of the  residual
interaction, $\Delta E_0$ (in KeV). In the last columns we report  the
amplitude with which the GDR's appear in such mixed states.}
\label{dcpb}
{\small

\begin{tabular}{||rcl|r||rr|r|r||} \hline
Dipole && States     &$    E_0 $(MeV)&$ \Delta E $&$ (\Delta E_0)  $
&$ c_{_{GDR_1}}$ &$ c_{_{GDR_2}}$\\ \hline \hline
$GDR_1\!\!$&$         $&$            $&$ 12.435 $&$ -132. $
	&$ (0.)    $&$ 0.993 $&$-0.006 $\\         
$GDR_2\!\!$&$         $&$            $&$ 16.662 $&$  -56. $
	&$ (0.)    $&$ 0.002 $&$ 0.994 $\\ \hline  
$3^-\!\!$&$\!\otimes\!$&$\!\!2^+     $&$  9.009 $&$  195. $
	&$ (200.)  $&$ 0.023 $&$ 0.000 $\\         
$3^-\!\!$&$\!\otimes\!$&$\!\!ISGQR   $&$ 15.062 $&$   75. $
	&$ (67.)   $&$ 0.045 $&$ 0.000 $\\         
$GDR_1\!\!$&$\!\otimes\!$&$\!\!2^+   $&$ 17.981 $&$ -207. $
	&$ (-220.) $&$ 0.043 $&$ 0.082 $\\         
$GDR_2\!\!$&$\!\otimes\!$&$\!\!2^+   $&$ 22.207 $&$  -23. $
	&$ (-36.)  $&$ 0.007 $&$ 0.048 $\\         
$GDR_1\!\!$&$\!\otimes\!$&$\!\!ISGQR $&$ 24.034 $&$   33. $
	&$ (-10.)  $&$ 0.057 $&$-0.004 $\\         
$3^-\!\!$&$\!\otimes\!$&$\!\!IVGQR   $&$ 25.278 $&$    6. $
	&$ (4.)    $&$-0.014 $&$ 0.000 $\\         
$GDR_1\!\!$&$\!\otimes\!$&$\!\!GMR_1 $&$ 26.046 $&$   18. $
	&$ (-27.)  $&$ 0.057 $&$ 0.000 $\\         
$HEOR\!\!$&$\!\otimes\!$&$\!\!2^+    $&$ 26.847 $&$   25. $
	&$ (24.)   $&$-0.004 $&$ 0.000 $\\         
$GDR_1\!\!$&$\!\otimes\!$&$\!\!GMR_2 $&$ 27.458 $&$  -10. $
	&$ (-35.)  $&$ 0.039 $&$ 0.000 $\\         
$GDR_2\!\!$&$\!\otimes\!$&$\!\!ISGQR $&$ 28.261 $&$  -88. $
	&$ (-51.)  $&$ 0.007 $&$ 0.054 $\\         
$HEOR\!\!$&$\!\otimes\!$&$\!\!ISGQR  $&$ 32.901 $&$  -31. $
	&$ (-30.)  $&$-0.004 $&$ 0.000 $\\         
$GDR_1\!\!$&$\!\otimes\!$&$\!\!IVGQR $&$ 34.250 $&$  -44. $
	&$ (-47.)  $&$-0.007 $&$ 0.000 $\\         
$GDR_2\!\!$&$\!\otimes\!$&$\!\!IVGQR $&$ 38.477 $&$ -174. $
	&$ (-174.) $&$ 0.006 $&$ 0.000 $\\         
$HEOR\!\!$&$\!\otimes\!$&$\!\!IVGQR  $&$ 43.117 $&$  -49. $
	&$ (-53.)  $&$-0.011 $&$ 0.000 $\\         
\hline \hline                                                          
\end{tabular}
}
\end{table}

Then for each spin and parity the total matrix has been diagonalised in
order to get the states $|\Phi_\alpha>$. Since these states are always
dominated by one component we have decided to label them by the name of
this dominant component. Table \ref{dcpb} gives for all $1^-$ states the
total shift $\Delta E$ (in KeV) from the unperturbed energy $E_0$ and
their component on the GDR. Tables \ref{lec} and \ref{hec} contain some
information on the results of the diagonalization for the low lying and
the high lying two-phonon states, respectively (see caption).   We have
restricted these tables to natural parity states  since the non natural
parity states are essentially not mixed and weakly excited. Moreover,
states with angular momentum greater than 3 have not been included
since they do not play an important role in Coulomb excitation
processes. From these tables one can see that the anharmonicities
predicted by our microscopic calculations are small, the typical shifts
in energy ($\Delta E$) being a few hundred keV. Each multiplet appears
to be splitted with a characteristic spreading equal to the global
shift. The mixing coefficients are in average also small, around 0.05
and at maximum around 0.2.

\begin {table} 
\caption { As table \ref{dcpb}, but for the low lying states  with
natural parity. In the first column we give the dominant component while
in the last one  we report the second most important component and its
coefficient.}
\label{lec}
{\small
\begin{tabular}{||rcl|r||c|rr|rrcl||} \hline
States&&&$    E_0 $(MeV)&$  J^\pi $&$ \Delta E $&$ (\Delta E_0)  $
&$c_{conf}$&Config.&& \\ \hline \hline
$ 3^-$&&&$  3.464 $&$ 3^- $&$ -256. $&$ (0.)    $&$  0.093$
	&$ 3^-\!\!$&$\otimes$&$\!\!2^+ $\\ \hline
$ 2^+$&&&$  5.545 $&$ 2^+ $&$-364.  $&$ (0.)    $&$ -0.201$
	&$ 3^-\!\!$&$\otimes$&$\!\!3^- $\\ \hline  \hline  
$3^-\!\!$&$\!\otimes\!$&$\!\!3^-  $&$  6.927 $&$ 0^+ $&$ 958. $
	&$ (1137.)$&$-0.163 $&$ GMR_1\!\!$&$ $&$ $\\
$ $&$  $&$ $&$                               $&$ 2^+ $&$ 381. $
	&$ (400.) $&$ 0.195 $&$ 2^+\!\!  $&$  $&$   $\\ \hline  
$3^-\!\!$&$\!\otimes\!$&$ 2^+\!\! $&$  9.009 $&$ 1^- $&$ 195. $
	&$ (200.) $&$-0.024 $&$ GDR_1\!  $&$ $&$ $\\
$ $&$  $&$ $&$                               $&$ 3^- $&$ 161. $
	&$ (112.) $&$-0.091 $&$ 3^-\!\!  $&$ $&$ $\\ \hline  
$2^+\!\!$&$\!\otimes\!$&$ 2^+\!\! $&$ 11.090 $&$ 0^+ $&$ 136. $
	&$ (145.) $&$-0.055 $&$ 3^-\!\!  $&$\otimes$&$\!\!3^- $\\
$ $&$  $&$ $&$                               $&$ 2^+ $&$ 178. $
	&$ (30.)  $&$-0.158 $&$ 2^+\!\!  $&$ $&$ $\\ \hline
$GDR_1\!\!$&$\!\otimes\!$&$2^+\!\!$&$ 17.981 $&$ 1^- $&$-207. $
	&$ (-220.)$&$-0.083 $&$ GDR_2\!  $&$ $&$ $\\
$ $&$  $&$ $&$                               $&$ 3^- $&$  -4. $
	&$ (-4.)  $&$-0.010 $&$ GMR_1\!\!$&$\otimes $&$ 3^- $\\ \hline  
\hline \hline
\end{tabular}
}
\end{table}

\begin {table} 
\caption { Same as table \ref{lec}, but for mixed states with natural
parity and with energies  between 22 and 29 MeV. }
\label{hec}

{\footnotesize
\begin{tabular}{||rcl|c||c|rr|rrcl||} \hline
States   &$ $&$    $&$    E_0 $(MeV)&$  J^\pi $&$ \Delta E  $
&$ (\Delta E_0)  $& $c_{conf}$&Config.&&\\ \hline \hline 
$ GDR_2\!\!$&$\otimes$&$\!\!2^+  $&$ 22.207 $&$ 1^- $&$ -23.$
  &$ (-36.) $&$ -0.046 $&$ GDR_2\!\!$&$ $&$ $\\
$ $&$    $&$     $&$                        $&$ 3^- $&$ -66.$
  &$ (-64.) $&$ -0.018 $&$ GDR_2\!\!$&$\otimes $&$ISGQR $\\ \hline  
$ ISGQR\!\!$&$\otimes$&$\!\!ISGQR$&$ 23.198 $&$ 0^+ $&$   4.$
  &$ (3.)   $&$  0.014 $&$ GDR_1\!\!$&$\otimes$&$\!\!GDR_1 $\\
$ $&$    $&$     $&$                        $&$ 2^+ $&$  35.$
  &$ (-15.) $&$ -0.061 $&$ ISGQR\!\!$&$ $&$ $\\ \hline  
$ GDR_1\!\!$&$\otimes$&$\!\!ISGQR$&$ 24.034 $&$ 1^- $&$  33.$
  &$ (-10.) $&$ -0.057 $&$ GDR_1\!\!$&$ $&$ $ \\
$ $&$  $&$       $&$                        $&$ 3^- $&$  -2.$
  &$ (-2.)  $&$  0.010 $&$ 3^-\!\!$&$\otimes$&$\!\!IVGQR $\\ \hline
$ 3^-\!\!$&$\otimes$&$\!\!HEOR   $&$ 24.766 $&$ 0^+ $&$  34.$
  &$ (14.)  $&$  0.133 $&$ GDR_1\!\!$&$\otimes$&$\!\!GDR_1 $\\ 
$ $&$  $&$       $&$                        $&$ 2^+ $&$  22.$
  &$ (-2.)  $&$  0.225 $&$ GDR_1\!\!$&$\otimes$&$\!\!GDR_1 $\\ \hline  
$ GDR_1\!\!$&$\otimes$&$\!\!GDR_1$&$ 24.871 $&$ 0^+ $&$  41.$
  &$ (33.)  $&$ -0.132 $&$ 3^-\!\!$&$\otimes$&$\!\!HEOR $\\
$ $&$  $&$       $&$                        $&$ 2^+ $&$-189.$
  &$ (-192.)$&$ -0.223 $&$ 3^-\!\!$&$\otimes$&$\!\!HEOR  $\\ \hline 
$ GMR_1\!\!$&$\otimes$&$\!\!ISGQR$&$ 25.210 $&$ 2^+ $&$  42.$
  &$ (11.)  $&$ -0.048 $&$ ISGQR\!\!$&$ $&$  $\\ \hline 
$ IVGQR\!\!$&$\otimes$&$\!\!3^-  $&$ 25.278 $&$ 1^- $&$   6.$
  &$ (4.)   $&$  0.015 $&$ GDR_1\!\!$&$ $&$  $\\
$ $&$   $&$      $&$                        $&$ 3^- $&$ -24.$
  &$ (-25.) $&$ -0.018 $&$ HEOR\!\!$&$ $&$  $\\ \hline  
$ GMR_1\!\!$&$\otimes$&$\!\!GDR_1$&$ 26.046 $&$ 1^- $&$  18.$
  &$ (-27.) $&$ -0.057 $&$ GDR_1\!\!$&$ $&$ $\\ \hline
$ GMR_2\!\!$&$\otimes$&$\!\!ISGQR$&$ 26.621 $&$ 2^+ $&$  25.$
  &$ (8.)   $&$ -0.033 $&$ ISGQR\!\!$&$ $&$  $\\ \hline 
$ 2^+\!\!$&$\otimes$&$\!\!HEOR   $&$ 26.847 $&$ 1^- $&$  25.$
  &$ (24.)  $&$  0.007 $&$ GMR_2\!\!$&$\otimes$&$\!\!GDR_1 $\\
$ $&$  $&$       $&$                        $&$ 3^- $&$ -39.$
  &$ (-44.) $&$ -0.018 $&$ HEOR\!\!$&$ $&$ $\\ \hline  
$ GMR_1\!\!$&$\otimes$&$\!\!GMR_1$&$ 27.221 $&$ 0^+ $&$ 297.$
  &$ (56.)  $&$ -0.128 $&$ GMR_1\!\!$&$ $&$ $\\ \hline 
$ 2^+\!\!$&$\otimes$&$\!\!IVGQR  $&$ 27.360 $&$ 0^+ $&$-196.$
  &$ (-195.)$&$ -0.016 $&$ ISGQR\!\!$&$\otimes$&$\!\!IVGQR $\\
$ $&$  $&$       $&$                        $&$ 2^+ $&$ -84.$
  &$ (-90.) $&$ -0.032 $&$ IVGQR\!\!$&$ $&$ $\\ \hline  
$ GMR_2\!\!$&$\otimes$&$\!\!GDR_1$&$ 27.458 $&$ 1^- $&$ -10.$
  &$ (-35.) $&$ -0.042 $&$ GDR_1\!\!$&$ $&$ $\\ \hline
$ GDR_2\!\!$&$\otimes$&$\!\!ISGQR$&$ 28.261 $&$ 1^- $&$  88.$
  &$ (51.)  $&$ -0.055 $&$ GDR_2\!\!$&$ $&$ $ \\
$ $&$  $&$       $&$                        $&$ 3^- $&$ -60.$
  &$ (-62.) $&$  0.010 $&$ 2^+\!\!$&$\otimes$&$\!\!GDR_2 $\\ \hline
$ GMR_1\!\!$&$\otimes$&$\!\!GMR_2$&$ 28.633 $&$ 0^+ $&$ 254.$
  &$ (74.)  $&$ -0.100 $&$ GMR_2\!\!$&$\otimes$&$\!\!GMR_2  $\\ \hline 
$ GDR_1\!\!$&$\otimes$&$\!\!GDR_2$&$ 29.097 $&$ 0^+ $&$-178.$
  &$ (-182.)$&$ -0.014 $&$ 2^+\!\!$&$\otimes$&$\!\!ISGQR$\\
$ $&$  $&$       $&$                        $&$ 2^+ $&$ -64.$
  &$ (-65.) $&$ -0.009 $&$ GDR_1\!\!$&$\otimes$&$\!\!GDR_1  $\\ \hline 
\end{tabular}
}

\end{table}

\subsection{ Excitation Processes }

Let us now study the characteristics of the excitation strength. We have
seen that the excitation operator contains 3 parts. The first one is the
linear response which is usually taken into account in the standard
calculations: i.e. in the harmonic and linear picture.  The strength
associated to the operator $W^{10}$ is, in this picture, concentrated in
the one-phonon states. The introduction of a mixing between states with
different numbers of phonons spreads the strength over more states. For
instance, in the case of the GDR the strength will be distributed among
the dipole states of tables \ref{dcpb} in a fashion
proportional to the c's coefficients. Analogously, the strength $W^{20}$
initially located around the two-phonon states, after the
diagonalisation will be distributed over many states. 

Moreover, the various states have now two paths to be excited in one
step, either through the $W^{10}$ excitation of their one-phonon
component or via the $W^{20}$ interaction exciting directly their
two-phonon part. Now, depending on the respective sign of the mixing
coefficients, these two contributions may interfere constructively or
destructively.

In addition to these direct transitions from the ground-state, the term
$W^{11}$ of the external field may induce transitions  between excited
states. These new excitation routes may modify the distribution of the
excitation probabilities associated with different states. In the next
subsection we will give a few examples where we will show the importance
of the $W^{11}$ and $W^{20}$ terms and of the anharmonicities.

\subsection{Excitation cross-sections}

Let us now put all these ingredients together in order to compute the 
excitation probabilities and cross sections. All the natural parity
states with angular momentum less or equal to 3 have been included in
the calculations while for the non natural parity states we have
included only the $1^+$ and the $2^-$ ones.  By solving the coupled
equations (\ref{adot}) we get the probability amplitude for each 
$|\phi_\alpha>$ state, from which we calculate the cross section by
integrating over the various impact parameters associated with Coulomb
inelastic excitations.  The $b_{min}$ has been chosen according to the
systematics of ref \cite{Be89}. We will describe in detail the results
for the $^{208}$Pb$ + ^{208}$Pb system at 641 MeV/A and we will first
focus our discussion on the excitation of dipole states.

\begin{figure} 
\begin{center}
\mbox{{\epsfxsize=13truecm \epsfysize=19truecm \epsfbox{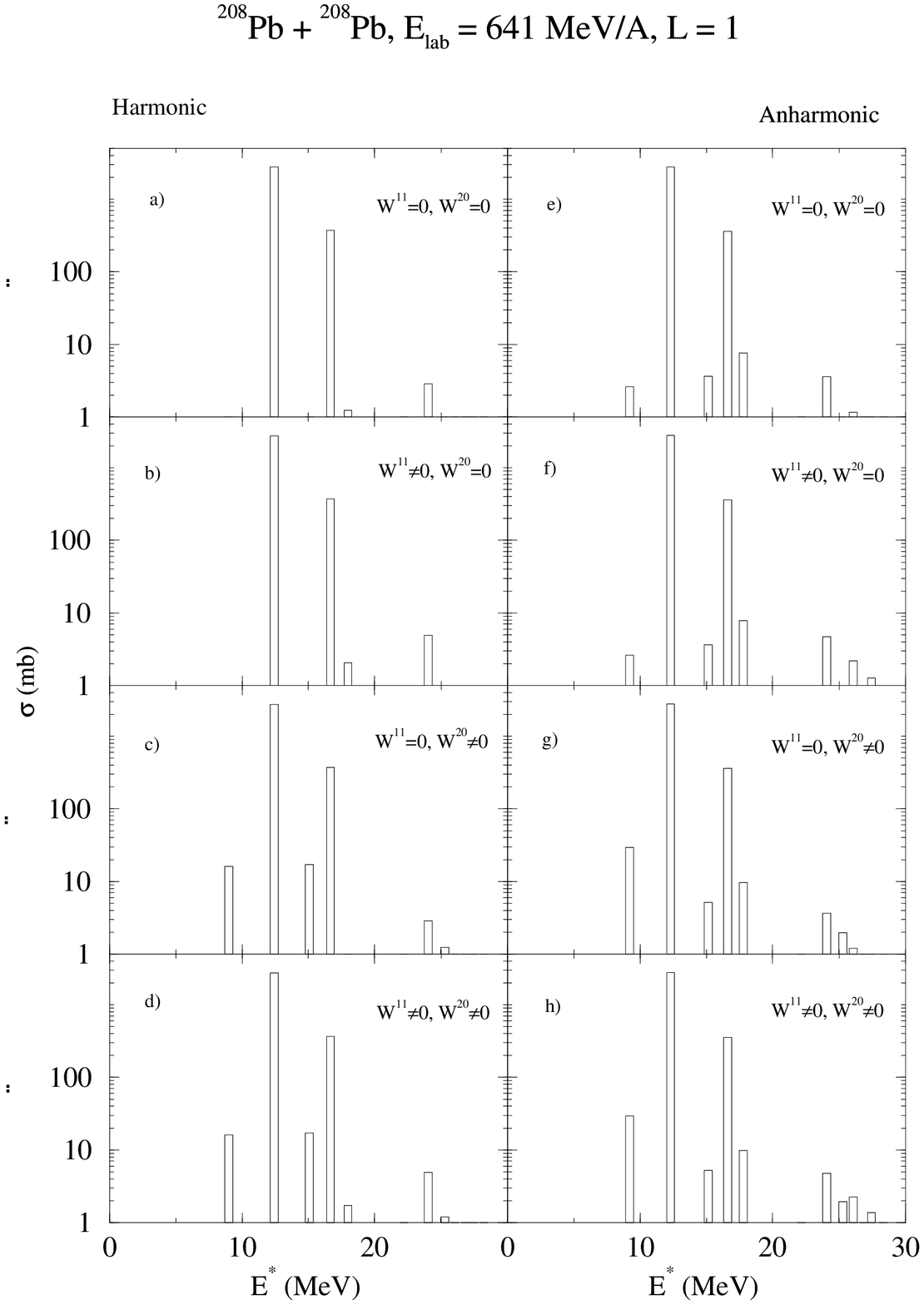}}}
\end{center}

\caption {Relativistic Coulomb target dipole excitation cross section 
for the $^{208}$Pb$ + ^{208}$Pb system at 641 MeV/A.
Each bar corresponds to the cross section of a single state. } 
\label{l1}
\end{figure}

In fig. \ref{l1} we present the dipole excitation cross-section as
predicted using various approximations in order to disentangle the
effects of the anharmonicities and non-linearities coming from $W^{11}$
and  $W^{20}$. We have run several calculations corresponding to the
various cases we can have, by switching on and off the different terms
of the external field. From the figure it is clear that the spectrum is
dominated by the dipole resonance. However, one can observe  important
modifications of the dipole strength for the different calculations
compared with the harmonic and linear prediction. 

\begin{figure} 
\begin{center}
\mbox{{\epsfxsize=11truecm \epsfysize=9truecm \epsfbox{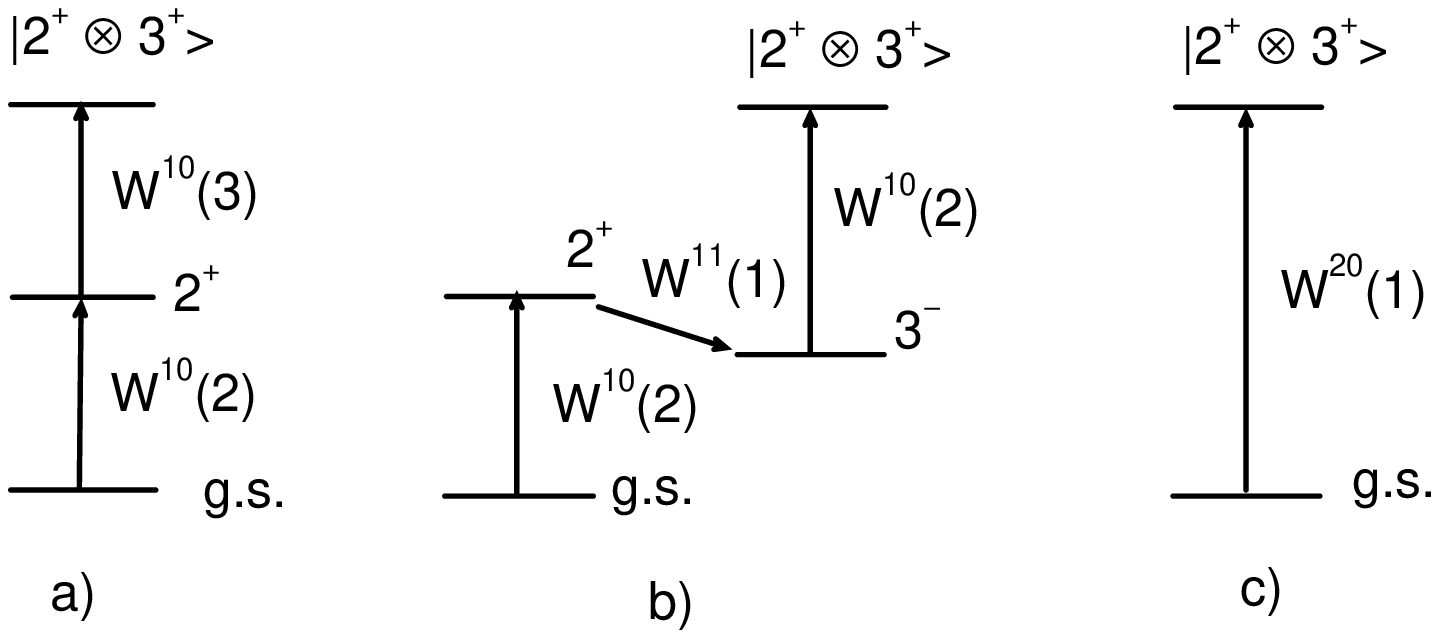}}}
\end{center}

\caption {Schematic representation of the Coulomb excitation of the 
$|2^+ \otimes 3^->$ state. }
\label{gs-23}
\end{figure}

In particular, states which were not excited in the harmonic and linear
picture can reach a sizeable cross-section when all the different
corrections are taken into account.  For instance, this is the case for
the state around 9 MeV, which is mostly built out of the 1$^-$ component
of the states resulting from the coupling of the low-lying 3$^-$ and
2$^+$.  In the first line of table \ref{spb}, the Coulomb inelastic
cross-sections for this state at several degrees of approximation are
given. One can see that this two-phonon state is almost not excited in
the harmonic and linear picture. Indeed, at this level of approximation,
the most direct way to excite this state requires one E$3$ and one E$2$ 
transitions (see figure \ref{gs-23}.a) which are not favourable. In this
case the $W^{11}$ term does not help much because either we reach the
state by one E$1$ plus two E$2$ transitions, as in figure \ref{gs-23}.b,
or by one E$3$ plus two E$1$ if in the first step we excite the $3^-$
state. In any case, at least one of the involved transitions  is of high
multipolarity. Conversely, the direct transitions due to the $W^{20}$
terms (see fig. \ref{gs-23}.c) increases the cross section by a huge
factor, bigger than 500. Indeed, this term is now a dipole transition
which is strongly favoured.  The importance of $W^{20}$ will decrease as
the excitation energy of the state increases. For instance, the
enhancement factor 500  reduces to about 50 for the dipole states
$|2^{+} \otimes HEOR>$ or $|ISGQR \otimes HEOR>$ whose energies are
around 30 MeV. 

\begin {table} 
\caption { Coulomb inelastic target excitation cross sections (in mb)
for the $^{208}$Pb$ + ^{208}$Pb system at 641 MeV/A and for  the mixed
states which are identified by their dominant component (first column)
and their angular momenta and parity (second column). In the third
column is  shown the reference result corresponding to a harmonic and
linear calculation. In the fourth column the additional inclusion of
only the $W^{11}$ non-linear  term is allowed.  Similarly, in the fifth
column the only difference with the reference calculation is due to the
addition of only the $W^{20}$ non-linear term.  In the sixth column the
results of an anharmonic and linear calculation are presented. The last
column correspond to results of the anharmonic and non-linear approach.}
\label{spb}

{\small
\begin{tabular}{||rcl|c||r||r|r|r||r||} \hline \hline
 States &&  &$J^\pi$&    harm. &$  W^{11} $&$ W^{20} $
& anharm.&anharm. \\
&&&&\& lin. &&&&\& non-lin. \\ \hline  \hline
$ 2^+   $&$ \otimes$&$  3^-      $&$ 1^- $&$  0.03 $&$  0.04 $
	&$ 16.21 $&$ 2.60  $&$  29.53 $\\  \hline
$ ISGQR $&$ \otimes$&$  3^-      $&$ 1^- $&$  0.05 $&$  0.07 $
	&$ 17.22 $&$ 3.63  $&$   5.18 $\\  \hline
$ 22 <  $&$ E     $&$ < 28 $ (MeV)&$ 1^- $&$  3.55 $&$  5.95 $
	&$  5.07 $&$ 6.42  $&$  12.18 $\\ \hline
$ 2^+   $&$ \otimes$&$  GDR_1    $&$ 1^- $&$  1.24 $&$  2.07 $
	&$  0.99 $&$ 7.64  $&$   9.83 $\\  \hline
 \hline \hline                                      
$ ISGQR $&$        $&$           $&$ 2^+ $&$298.91 $&$332.56 $
	&$300.09 $&$278.35 $&$ 314.18 $\\  \hline \hline
\end{tabular}

}
\end{table}

When the mixing of one- and two-phonons states is taken into account 
this state can be also populated by $W^{10}$ through its small GDR
component  (see table \ref{dcpb}). In fact, although the c coefficient
of the GDR component is small, this component gives a considerable
contribution due to  the fact that it is a one step  dipole excitation.
Moreover, the energy of the state (about 9 MeV) is lower than the  one
of the GDR state. All together  the effect of the anharmonicities on the
inelastic cross section is a factor about 100 times bigger with respect
to our reference calculation.

Finally, when all these different contributions are taken into account
this dipole two-phonon state built from low-lying 3$^-$ and 2$^+$ is
receiving  30 mb cross section, while in the harmonic and linear limit
it was just 0.03 mb. 

In this case the effects of non-linearities and anharmonicities
interfere constructively. That is not a general property. An example  in
which these effects interfere destructively is shown in table \ref{spb}, 
where the excitation cross section to the dipole state  $|ISGQR \otimes
3^{-}>$ is given. In order to clarify this mechanism we have done a
parametric calculation in which only three single phonon states were
considered, namely the 3$^-$, the 2$^+$ and the GDR.  Then we have mixed
the single phonon $|GDR>$ with the two phonon state $|2^{+} \otimes
3^{-}>$, coupled to a total spin 1, in the following way
\begin{eqnarray}
|\Phi_1> &=& \cos \beta |2^{+} \otimes 3^{-}> + \sin \beta |GDR>  
\nonumber\\
|\Phi_2> &=& -\sin \beta |2^{+} \otimes 3^{-}> + \cos \beta |GDR> 
\label {para}
\end{eqnarray}
\noindent Increasing the parameter $|\beta|$ we can go from a pure
harmonic case ($\beta=0$) to a very strong anharmonicity. Changing the
$\beta$ sign the relative phases of the $|\Phi_{\alpha}>$ components are
changed.  The energies of the states were kept fixed and equal to the
energy, in the harmonic limit, of the main component; i.e.  $E_1=
E_{2^+} + E_{3^-}$ and $E_2= E_{GDR}$. 

\begin {table} 
\caption { Same as table \ref{spb}, but for the parametric state
$|\Phi_1>$ of eq. (\ref{para}). In the last column the values of the
parameter $\beta$ used.}
\label{tbeta}

\begin{tabular}{||c|c|c|c|c|r||} \hline \hline
    harm. \& lin. &$  W^{11} $&$ W^{20} $& anharm.
&anharm. \& non-lin. & $\sin \beta$ \\ \hline  \hline
$      $&$      $&$      $&$ 1.96 $&$ 29.71 $&$ -0.02$\\  
$ 0.26 $&$ 0.27 $&$ 16.58$&$      $&$       $&\\  
$      $&$      $&$      $&$ 1.92 $&$  7.20 $&$  0.02$\\ 
 \hline \hline
\end{tabular}

\end{table}

The cross sections corresponding to the $|\Phi_1>$ state are shown in
table \ref{tbeta} for two opposite values of $\beta$. From the table, we
can see that the behaviour of the cross section is very similar to the
one obtained in the complete calculation (see table \ref{spb}). Indeed
the results for $\beta=-0.02$ are similar to the ones obtained for the
$|2^{+} \otimes 3^{-}>$ dipole state,  where anharmonicities reinforce
the effects of non-linearities on the cross section. Conversely, for
$\beta=0.02$ the final result is much lower than the one given by the
$W^{20}$ term alone.  This result is very similar to the one obtained
for the  $|ISGQR \otimes 3^{-}>$ state shown in table \ref{spb}.

The reason for this different behaviour can be easily understood  in a
first order calculation if we take into account the following relations 
\begin{eqnarray}
<\nu \lambda \mu| W(t)|0> &=& 
g_{E \lambda \mu}(\beta,t) <\nu \lambda|V^{10}(E\lambda)|0> \nonumber\\
<\left[\nu_1 \nu_2 \right] \lambda \mu| W(t)|0> &=&
{1 \over {\sqrt{1+\delta_{\nu_1,\nu_2}}}} 
g_{E \lambda \mu}(\beta,t) 
<\nu_1 \lambda_1 \nu_2 \lambda_2|V^{20}(E\lambda)|0> 
\end{eqnarray}
\noindent where $g_{E \lambda \mu}$ was defined in equation
(\ref{multi}).  Let us call  $\sigma_1$ the cross section corresponding
to the state $|\Phi_1>$. In a first order calculation we get 
\begin{equation}\label{38}
{\sigma_1^{{\rm anharm} \& {\rm non-lin}} \over 
 \sigma_1^{{\rm harm} \& {\rm non-lin}}} = 
\left( \cos \beta + {{\sin \beta} \over x} \right)^2
\end{equation}
\noindent where $x$ is given by the following ratio of matrix elements
\begin{equation}
x= {{<2^{+} \otimes 3^{-}|V^{20}(E1)|0>} \over {<GDR|V^{10}(E1)|0>}}
\end{equation}
\noindent Since $|x|$ is usually smaller than 1 the second term in
equation (\ref{38}) can be important even for small anharmonicities.
The values of $\beta$ and $x$ as well as their signs are important.

In the same way we can calculate the cross section $\sigma_2$
corresponding to the state $|\Phi_2>$ and get

\begin{equation}
{\sigma_2^{{\rm anharm} \& {\rm non-lin}} \over 
 \sigma_2^{{\rm harm} \& {\rm non-lin}}} = 
\left( \cos \beta - x \enskip{\sin \beta} \right)^2
\end{equation}

\noindent Since $|x|$ is usually small, the previous ratio will not
differ very much from one. Note also that, in first order,
$\sigma_2^{{\rm harm} \& {\rm non-lin}}$  and  $\sigma_2^{{\rm harm} \&
{\rm lin}}$ coincide, while  $\sigma_1^{{\rm harm} \& {\rm non-lin}}$
differs from $\sigma_1^{{\rm harm} \& {\rm lin}}=0$. 

\begin{figure} 
\begin{center}
\mbox{{\epsfxsize=9truecm \epsfysize=9truecm \epsfbox{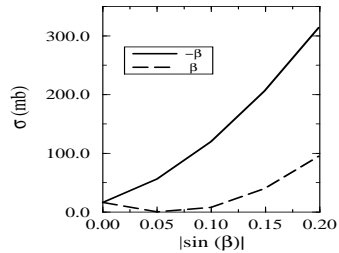}}}
\end{center}

\caption { Relativistic Coulomb target excitation cross section for the
parametric calculation of eq. (\ref{para}) as function of absolute value
of the mixing coefficient $\sin \beta$. } 
\label{beta}
\end{figure}

Now, it happens that the $x$ ratio for the $|2^{+} \otimes 3^{-}>$
and $|ISGQR \otimes 3^{-}>$ dipole states has the same sign and similar
values: $-$0.058 and $-$0.091, respectively. But the coefficients of their
GDR component have opposite sign (see table \ref{dcpb}) and their values
are such that the dependence of the ratio in equation (\ref{38}) on
$\beta$ is nearly linear. Then in one case anharmonicities and
non-linearities interfere constructively and in the other case interfere
destructively. By increasing $|\beta |$ this property is lost and we
could have a reinforce effect in $\sigma_1$ even if $\beta$ and $x$ have
opposite sign. That is confirmed by the parametric calculation to all
orders, as can be seen in figure \ref{beta}, where we have increased the
$|\beta|$ parameter up to about 0.2. That would support that for nuclei
with stronger anharmonicities we should get a higher increase in the
cross section with respect to the linear and harmonic case.

Of course in this example we have assumed that the mixing has just two 
components what is a great simplification, specially for the state whose
main component is the GDR. Let us go back to the complete
calculations where we have a mixing of all the states and their proper
energies are taken into account.

Similar effects can also be seen on other states in the dipole response
of the Pb nucleus (see table \ref{hec}).  The two-phonon states located 
around 25 MeV excitation energy are of particular interest. These states
are mainly built by coupling the giant  dipole with the monopole and
quadrupole states. As in the previous case, the  direct transition
$W^{20}$ and the mixing are important. In addition to that, also the
transition between one-phonon states contributes to increase the cross
section. Table  \ref{spb} shows that in such a case the increase of the
cross section of these two-phonon states is more than 300\%. This
reminds the findings discussed in ref \cite{vol95}, where in a very
schematic model we were showing that non-linearities and anharmonicities
might strongly modify the excitation cross-section. An example where the
anharmonicities play an important role is given by the excitation of the 
$|GDR_1 \otimes 2^+>$ state whose cross section is reported in table
\ref{spb}. The big increase of the anharmonic and non-linear cross
section can be entirely ascribed to its big $GDR_2$ component (see table
\ref{lec}).

\begin{figure} 
\begin{center}
\mbox{{\epsfxsize=11truecm \epsfysize=9truecm \epsfbox{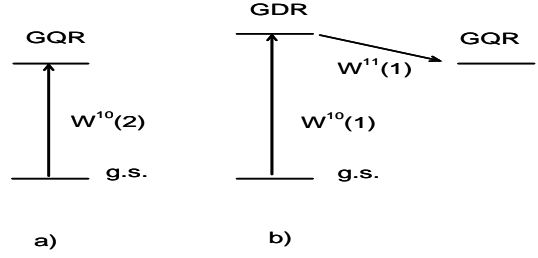}}}
\end{center}

\caption {Schematic representation of the Coulomb excitation of the 
$|GQR>$ state. }
\label{gs-gqr}
\end{figure}

This increase of the cross section is seen not only in the dipole
channel, which gets large  contributions from the GDR itself, but also
for other multipolarities. Let us for instance consider  the isoscalar
giant quadrupole resonance (ISGQR) (see table \ref{spb}). Looking at its
excitation we see that  the inclusion of $W^{11}$ raises the value of
$\sigma$ from 299 mb to  333 mb (in the harmonic case). In this case,
besides the direct  transition to the GQR due to the action of $W^{10}$,
we are considering  the second order one which proceeds first through
the excitation of the GDR by $W^{10}(1)$ and  then to GQR  by means of
$W^{11}$ (see fig. \ref{gs-gqr}).  This second order process is able to 
give almost a 12\% increase because the excitation probability  of the
first transition is very high and because the effect of $W^{11}$ is
enhanced by the fact that the energies of GDR and GQR are close each
other.

We close this detailed analysis with a comment on the effect of the non
natural parity states $1^+$ and $2^-$ we have introduced in the
calculations.  In our calculation we found that the contribution of the
$1^+$ states is, in this respect, irrelevant and the one of the $2^-$,
in the region of the DGDR, amounts to about 1 mb. At lower energy,
around 16 MeV, its contribution is 2 mb. Similar conclusions have been
recently reached in ref.\cite{be96}.

\begin{figure} 
\begin{center}
\mbox{{\epsfxsize=13truecm \epsfysize=9truecm \epsfbox{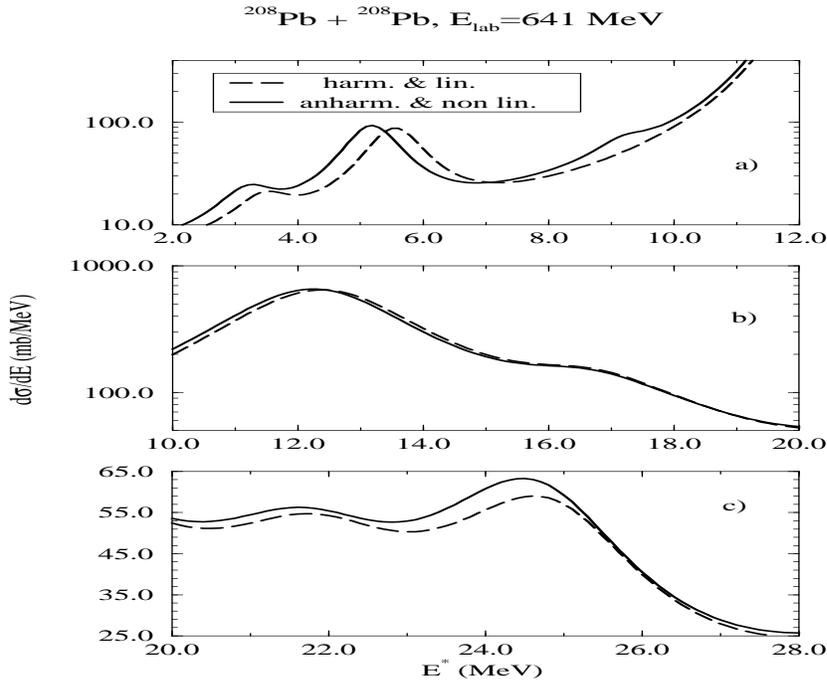}}}
\end{center}

\caption {Relativistic Coulomb target excitation  cross section for the
$^{208}$Pb$ + ^{208}$Pb system at 641 MeV/A as function of the
excitation energy. The three parts correspond to different energy
regions. The cross section for each $|\Phi_\alpha>$ state has been
smoothed by a lorentzian with a 3 MeV width. For the low energy region we
used a 1 MeV width.} 
\label{dsde1}
\end{figure}

So far, we have discussed the influence on some particular states. In
order to get a global view on the effects of both non-linearities and
anharmonicities  we must compute the complete inelastic cross section.
Therefore, we have summed up all the contributions coming from the
various states after a smoothing of each individual line shape by a
lorentzian. The results are presented in fig. \ref{dsde1}. For the low
energy region (fig. \ref{dsde1}.a) the width of the lorentzian has been
chosen equal to 1 MeV, while for the energy region around the GDR (fig.
\ref{dsde1}.b) and the one around the DGDR (fig. \ref{dsde1}.c) it has
been fixed equal to 3 MeV. In this figure we can see that the single GDR
region is not much affected by the anharmonicities and non-linearities
while the cross-section in the DGDR region is increased by 10\%
when the anharmonicities and non-linearities are taken into account. We
would like to point out that this increase is mainly due to the
excitation of two-phonon states whose energies are in the DGDR
region and whose population has been possible only because of the
presence of the anharmonicities and the non-linear terms $W^{11}$ and
$W^{20}$ in the external field. The low lying part of the spectrum is
also affected and in particular, as we discussed before, a new dipole
strength is visible in the 9 MeV region.

\begin {table} 
\caption { Comparison between our theoretical results and the
experimental cross sections (in barn) reported in ref. [6] 
for the Pb + Pb  reaction at 641 MeV per nucleon.  The theoretical
results (first line) correspond to the sum of all GDR (first column) and
all DGDR (second column)  cross-section.  The third column contains the
cross section associated with all the states above the IVGQR (E $>$ 22
MeV). The theoretical cross sections are obtained from the non-linear
and anharmonic  calculation while the numbers in parenthesis refer to
the linear and harmonic  limit. The experimental results are reported in
the second line.  The first number corresponds to the extracted GDR
cross section while the second number comes from a gaussian fit of the
high energy cross section  after subtraction of the GDR and GQR
single-phonon strength. }
\label{tsigma}

\begin{tabular}{||c||c||c|c||} \hline \hline
                  & GDR  & DGDR & DGDR energy region
\\ \hline  \hline
$\sigma_{\rm th}$&$3.13$ $(3.14)$&$0.21$ $(0.22)$&$0.31$ $(0.28)$ 
		\\ \hline  
$\sigma_{\rm exp}$ & $3.28 \pm 0.05$ & \multicolumn{2}{c||}
       		{$0.38\pm 0.04$} \\   \hline \hline
\end{tabular}

\end{table}

In table \ref{tsigma} we show a comparison between our theoretical
results and the experimental cross-section for the GDR and the DGDR
energy region. The agreement for the GDR seems satisfactory.  The
theoretical yield associated with the DGDR states  explains about 60\%
of the experimental cross section.  However, this disagreement between
the experimental cross section in the DGDR region  and our theoretical
estimate is reduced to 18\% $\pm$ 10\% by the inclusion of all  the
different multiphonon states considered in our calculation and lying
above the IVGQR.

In conclusion,  both the introduction of different two-phonon states and 
the inclusion of anharmonicities and non-linearities are bringing the 
theoretical prediction rather close to the experimental observation for
the Coulomb excitation of Pb nuclei in the DGDR region.

\section{Results about the excitation of $^{40}$Ca}

We have also done calculations for the excitation of $^{40}$Ca by a 
$^{208}$Pb projectile with $E_{lab}=1000 MeV/A$. The one-phonon basis
for  $^{40}$Ca are shown in table \ref{fca}. We do not have any
collective low lying $2^+$ state because the RPA does not generate
such state for the $^{40}$Ca nucleus. The properties of the dipole $1^-$
states are reported in  table \ref{dcca}, which is the analogous of
table \ref{dcpb} for $^{208}$Pb. We note that we have bigger
anharmonicities than in the Pb case. 

\begin {table} 
\caption { Same as table \ref{fpb}, but for $^{40}$Ca.}
\label{fca}

\begin{tabular}{||r||lr|r|r||} \hline
    Phonons   &$J^\pi$&T    &$E (MeV) $ & \% EWSR \\ \hline \hline
$GMR_1$& $0^+$&0 & 18.25 & 30 \\
$GMR_2$& $0^+$&0 & 22.47 & 54 \\ \hline
$GDR_1$& $1^-$&1 & 17.78 & 56 \\
$GDR_2$& $1^-$&1 & 22.03 & 10 \\ \hline
$ISGQR$& $2^+$&0 & 16.91 & 85 \\
$IVGQR$& $2^+$&1 & 29.53 & 26 \\ \hline 
$3^-  $& $3^-$&0 &  4.94 & 14 \\
$LEOR $& $3^-$&0 &  9.71 &  5 \\
$HEOR $& $3^-$&0 & 31.33 & 25 \\ \hline
\end{tabular}

\end{table}

\begin {table} 
\caption { Same as table \ref{dcpb}, but for $^{40}$Ca.}
\label{dcca}

{\small
\begin{tabular}{||rcl|r||rr|r|r||} \hline
Dipole  &&States     &$    E_0 $(MeV)&$ \Delta E $&$ (\Delta E_0)$
&$ c_{GDR_1}$&$ c_{GDR_2}$ \\ \hline \hline
$ GDR_1 $&$       $&$       $&$ 17.780 $&$ -432. $&$    0. $
	&$  0.989$&$ -0.006$               \\
$ GDR_2 $&$       $&$       $&$ 22.034 $&$ -391. $&$    0. $
	&$  0.004$&$  0.990$                \\
\hline                                                                        
$ ISGQR\!\!$&$\otimes$&$\!\!3-    $&$ 21.851 $&$ 708. $&$ 713. $
	&$ 0.024$&$ 0.011$          \\
$ ISGQR\!\!$&$\otimes$&$\!\!LEOR  $&$ 26.616 $&$ 231. $&$ 224. $
	&$ 0.011$&$-0.011$          \\
$ IVGQR\!\!$&$\otimes$&$\!\!3^-   $&$ 34.541 $&$-125. $&$-128. $
	&$ 0.001$&$-0.020$            \\
$ GDR_1\!\!$&$\otimes$&$\!\!ISGQR $&$ 34.690 $&$ 139. $&$  35. $
	&$-0.063$&$-0.044$     \\
$ GMR_1\!\!$&$\otimes$&$\!\!GDR_1 $&$ 36.026 $&$-110. $&$-214. $
	&$-0.075$&$-0.004$ \\
$ GDR_2\!\!$&$\otimes$&$\!\!ISGQR $&$ 38.943 $&$ -21. $&$ -74. $
	&$-0.034$&$ 0.034$       \\
$ IVGQR\!\!$&$\otimes$&$\!\!LEOR  $&$ 39.305 $&$-245. $&$-245. $
	&$ 0.000$&$ 0.003$             \\
$ GMR_1\!\!$&$\otimes$&$\!\!GDR_2 $&$ 40.280 $&$-175. $&$-292. $
	&$ 0.011$&$-0.079$ \\
$ GMR_2\!\!$&$\otimes$&$\!\!GDR_1 $&$ 40.249 $&$   9. $&$-202. $
	&$-0.098$&$-0.005$   \\
$ GMR_2\!\!$&$\otimes$&$\!\!GDR_2 $&$ 44.502 $&$  20. $&$-194. $
	&$ 0.000$&$-0.098$   \\
$ GDR_1\!\!$&$\otimes$&$\!\!IVGQR $&$ 47.379 $&$-315. $&$-308. $
	&$-0.011$&$-0.003$     \\
$ ISGQR\!\!$&$\otimes$&$\!\!HEOR  $&$ 48.240 $&$ -13. $&$ -27. $
	&$ 0.000$&$ 0.005$           \\
$ GDR_2\!\!$&$\otimes$&$\!\!IVGQR $&$ 51.633 $&$-270. $&$-271. $
	&$ 0.001$&$ 0.001$        \\
$ IVGQR\!\!$&$\otimes$&$\!\!HEOR  $&$ 60.929 $&$-271. $&$-275. $
	&$-0.009$&$ 0.005$              \\
\hline \hline                                                  
\end{tabular}
}

\end{table}

\begin{figure} 
\begin{center}
\mbox{{\epsfxsize=13truecm \epsfysize=9truecm \epsfbox{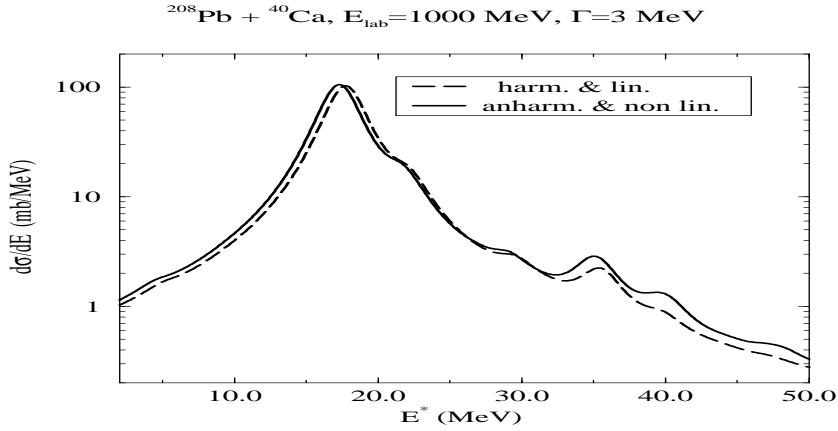}}}
\end{center}

\caption {Relativistic Coulomb target excitation  cross section for the
$^{208}$Pb$ + ^{40}$Ca system at 1000 MeV/A as function of the
excitation energy.  The cross section for each $|\Phi_\alpha>$ state has
been smoothed by a lorentzian with a 3 MeV width.} 
\label{ca}
\end{figure}

The coupled channel equations (\ref{adot}) were solved only for the
natural  parity states which, as we have seen in the case of Pb, are
providing the largest contribution to the cross-section. The resulting
cross section, after a smoothing by a lorentzian with a 3 MeV width, is
shown in fig. \ref{ca}. The peak at around 18 MeV is due to the $GDR_1$
with the contribution of the ISGQR state. The shoulder at about 22 MeV
is given by the $GDR_2$ and the two-phonon dipole state $|ISGQR \otimes
3^->$ which gives, in the anharmonic and non-linear case, a 10\%
increase. The latter state is excited in the same fashion of the $|ISGQR
\otimes 3^->$ state of $^{208}$Pb, see table \ref{sca}, with the 
difference that now the increasing factor is 1000 while in the
$^{208}$Pb case it was only 100. Finally, we note two interesting energy 
regions where there is a difference between the harmonic and linear and 
the anharmonic and non-linear case, namely the regions around 35 and 40
MeV. The sum of the cross section for the $1^-$ state belonging to these
two  regions are reported in table \ref{sca} for the different kinds of
calculations we can make within our approach. From the table we see that
the increase is essentially due to the dipole $1^-$ states and this is
an almost pure  anharmonic effect. The global increase in the DGDR
energy  region  amounts to a 20\%, which is twice what we obtained
for Pb. 

\begin {table} 
\caption { Same as table \ref{spb}, but for $^{40}$Ca.}
\label{sca}

{\small
\begin{tabular}{||rcl|c||r||r|r|r||r||} \hline \hline
 States &&  &$J^\pi$&    harm. &$  W^{11} $&$ W^{20} $
& anharm.&anharm. \\
&&&&\& lin. &&&&\& non-lin. \\ \hline  \hline
$ ISGQR $&$ \otimes$&$ 3^-        $&$ 1^-$&$ 0.004 $&$ 0.006$
	&$ 6.660$&$ 0.284 $&$ 3.955 $\\  \hline
$ 34 <  $&$ E      $&$ < 36 $ (MeV)&$ 1^-$&$ 0.110 $&$ 0.287$
	&$ 0.295$&$ 1.723 $&$ 2.221 $\\ \hline
$ 38 <  $&$ E      $&$ < 45 $ (MeV)&$ 1^-$&$ 0.008 $&$ 0.020$
	&$ 0.022$&$ 1.468 $&$ 1.698 $\\ \hline
 \hline \hline
\end{tabular}
}

\end{table}

\section{Discussion and conclusion}

We have employed an RPA based approach  to compute the anharmonicities: 
We have diagonalized the residual interaction between RPA phonons
in the space of one- and two-phonons states.  We have taken into
account also the  particle-particle and hole-hole terms in the external
field  making possible the direct excitation of two-phonon states as
well as the transition between one-phonon states. These non-linear terms
and the  anharmonicities are not taken into account in the "standard"
approach of  the multiphonon picture. Within this framework we have
calculated the Coulomb excitation of $^{208}$Pb and $^{40}$Ca nuclei due
to the impinging $^{208}$Pb nucleus at 641 and 1000 MeV/A.

In this paper we have shown that the inclusion of both anharmonicities
and non-linear terms in the external field reduce  the disagreement
between the experimental cross section  in the DGDR region and the
theoretical one calculated  within the "standard"  approach. Moreover,
for the  $^{208}$Pb case, we have found a big effect also at low energy
where the  $|2^+ \otimes 3^+>$ state would have never been excited
without the presence  of both the anharmonicities and non-linearities.
Since these low lying  two-phonon states are strongly  mixed and since
their energy is low, we  believe that they could also be strongly excited
by the nuclear part of the mean field at  an incident  energy lower than
the one considered here. Theoretical and  experimental work in this
direction are called for.

In view of our calculations it is clear that non-linearities and
anharmonicities have an influence on the Coulomb excitation. On some
particular states this influence can be very strong, while averaging
over all the states we have found an increase of the cross section by
about 10\% for Pb and 20\% for Ca in the region  of the two-phonon
states while the energies were modified only by a few percent. 
However, this might not be the final answer because of different
reasons. First, we are working in a truncated subspace in order to keep
only one- and two- phonon states. However, we know that a large part of
the increase observed in ref. \cite{vol95} is due to the increase of
transition matrix elements coming from components of the wave function
containing large phonon numbers. These components are not taken into
account in the present calculation and this reduces the influence of
anharmonicities. In fact we have tested this point on the simple model
reported in reference \cite{vol95} and we have observed that a
truncation of the multiphonon space at the two-phonon level  reduces the
increase of the cross section by almost a factor 4. Unfortunately, this
point is not easy to improve because the computation time will become
too long if we are forced to include more multiphonon states. We are now
trying to develop an alternative approach based on time-dependent mean
field theory in the boson representation.

The anharmonicities we are computing are mainly due to the
residual interaction into channels which are different from the usual
particle-hole interaction. One may argue that,  as far as effective
interactions are concerned, their parameters are only fitted close to
the ground state. Therefore, except  for the  particle-hole channel the
other parts of the interaction  are not really constrained by the
theory.  However, in some cases the residual interaction  has been
tested far from the ground state. In this respect the relative success
of the time-dependent mean field theory (and other treatments such as
adiabatic TDHF or generator coordinate method) might be an indication
that the same Skyrme parametrisation also holds for large amplitude
motion. However, this point is certainly calling for more theoretical
developments in order to better define the effective interaction in
channels different from the standard particle-hole ones.

From the experimental point of view it seems that the $^{208}$Pb nucleus
behaves as a rather good vibrator. In fact the discrepancy about the
cross-section between theory and experiment is apparently much smaller
in the case of Pb than in the case of Xe \cite{ST95}.  Moreover,  as far
as the shift in energy of the two-phonon state with respect to the
harmonic limit is concerned,  a shift of less than 1\% was found
experimentally for Pb while  for Xe it was of the order of 10\%
\cite{schm,ST95}. In our calculation the Pb also appears  as a rather
good vibrator and the predicted effects  on the energy shifts are
consistent with the experiment. This is probably related to the fact
that $^{208}$Pb is a double-magic nucleus. It would be very important to
study non-magic, open-shell or deformed nuclei which are expected to be
poorer vibrators than double-magic nuclei. In particular, it is known
that the energy of the GDR is strongly affected by the deformation,
indicating a possible strong coupling between dipole and quadrupole
degrees of freedom. This may induce a modification of the cross-section
stronger than the predicted 10 \% to 20\% for spherical-magic nuclei. In
this respect, extensions of the presented results to open shell and
deformed nuclei are called for.

In conclusion, we would like to stress that in addition to the DGDR
excitation  several states are contributing to the cross-section in the
DGDR energy region. When the non-linearities and the anharmonicities are
taken into account the total theoretical cross-section above the IVGQR
come rather close to the experimental result for the Coulomb excitation
of Pb.

\ack{ This work has been partially supported by the spanish DGICyT under
contract PB92-0663, by the Spanish-Italian agreement between the DGICyT
and the  INFN and by the Spanish-French agreement between the DGICyT and
the IN2P3. }

\section{ Appendix }

We just need to know

\begin{equation} \label{e3}
H_{\lambda \mu}(\beta,t)=\int_0^{+\infty} \big( e^{i\omega t}
(-1)^{\lambda +\mu}
+ e^{-i\omega t} \big) \omega^\lambda K_\mu (\beta\omega) d\omega
\end{equation}

\noindent
with $\mu \geq 0$, since  $H_{\lambda \mu}(\beta,t)=H_{\lambda
\vert\mu\vert}(\beta,t)$. Considering the cases $\lambda+\mu $ even or
odd, together with t positive, negative or null, the integral (\ref{e3})
will be proportional to  integrals in \cite{Gr}. Combining all cases we
get

\begin{eqnarray} \label{e4}
& &H_{\lambda \mu}(\beta,t) =    \nonumber \\
& &( 1+(-1)^{\lambda+\mu}) {2^{\lambda-1} \over \beta^{\lambda+1}}
\Gamma({\scriptstyle{{1+\lambda+\mu} \over 2}}) 
\Gamma({\scriptstyle{{1+\lambda-\mu} \over 2}})
F\big({\scriptstyle{
{{1+\lambda+\mu} \over 2},{{1+\lambda-\mu} \over 2};
{1\over 2};-{t^2\over \beta^ 2} }}) -    \nonumber \\
&i&( 1-(-1)^{\lambda+\mu}) {{2^\lambda t} \over \beta^{\lambda+2}} 
\Gamma({\scriptstyle{{2+\lambda+\mu} \over 2}}) 
\Gamma({\scriptstyle{{2+\lambda-\mu} \over 2}})
F\big({\scriptstyle{
{{2+\lambda+\mu} \over 2},{{2+\lambda-\mu} \over 2};
{3\over 2};-{t^2\over \beta^ 2} }})
\end{eqnarray}

\noindent These hypergeometric functions can be transformed following
\cite{Ab} as

\begin{eqnarray} \label{e5}
F(n+{\scriptstyle{1\over 2}},n+{\scriptstyle{1\over 2}}-\mu;
m+{\scriptstyle{1\over 2}}; -x^2) &=&
{1\over {(1+x^2)^{2n-\mu-m+1/2}}} \nonumber \\
&\times &F(m-n,m-n+\mu;m+{\scriptstyle{1\over 2}};-x^2)
\end{eqnarray}

If $\lambda+\mu$ is even, then $m=0$ and $n=\frac{\lambda+\mu}{2}$.
Whereas if $\lambda+\mu$ is odd, then $m=1$ and
$n=\frac{\lambda+\mu+1}{2}$. Therefore, in the cases in which we are
interested $m-n$ is an integer $\leq 0$, and the latter hypergeometric
function reduces to the following polynomial 

\begin{equation} \label{e6}
F(-\ell,b;c;z)= \sum_{k=0}^\ell {{(-\ell)_k(b)_k} \over {(c)_k}} 
{{z^k}\over {k!}} \> \>.
\end{equation}


\begin{thebibliography}{99}

\bibitem{bm}    A. Bohr and B.R. Mottelson, Nuclear Structure, vol. II, 
                (W.A. Benjamin, N.Y., 1975)
\bibitem{f88}   N. Frascaria, Nucl. Phys. A 282 (1988) 245c
\bibitem{mor88} S. Mordechai et al., Phys.Rev.Lett. 60(1988)408
\bibitem{schm}  R. Schmidt et al., Phys. Rev. Lett. 70 (1993) 1767 
\bibitem{ri93}  J. Ritman et al, Phys. Rev. Lett. 70 (1993) 533
\bibitem{ST95}  J. Stroth et al, in the Proceedings of the "Groningen 
                Conference on Giant Resonances", June 28-July 1, 1995, 
                Nucl. Phys. A 599 (1996)307c; K. Boretzky et al., GSI 
		preprint-96-27 to be published on Physics Letters B.
\bibitem{cho95} Ph. Chomaz and N. Frascaria, Phys. Rep. 252 (1995) 5
\bibitem{plb94} K.Govaert et al., Phys. Lett. B335(1994)113
\bibitem{cat89} F. Catara, Ph. Chomaz and N. Van Giai, 
                Phys. Lett. B 233 (1989) 6
\bibitem{bea92} D. Beaumel and Ph. Chomaz, Ann. Phys. (N.Y.) {\bf 213} 
                (1992) 405.
\bibitem{be96}  C. A. Bertulani, L. F. Canto, M. S. Hussein and 
                A. F. R. de Toledo Piza, Phys. Rev. C 53 (1996) 334 
\bibitem{vol95} C. Volpe, F. Catara, Ph. Chomaz, M. V. Andr\'es 
	  	and E. G. Lanza, Nucl. Phys. A 589 (1995) 521;
		Nucl. Phycs. A 599 (1996) 347c. 
\bibitem{rs}    P. Ring and P. Schuck, The nuclear many-body problem 
		(Springer, Berlin, 1981)
\bibitem{ca86}  F. Catara and U. Lombardo, Nucl. Phys. A 455 (1986) 158
\bibitem{ca87}  F. Catara, Ph. Chomaz and A. Vitturi, 
                Nucl. Phys. A 471 (1987) 661
\bibitem{ca88}  F. Catara and Ph. Chomaz, Nucl. Phys. A 482 (1988) 271c
\bibitem{ccg92} F. Catara, Ph. Chomaz and N. Van Giai, 
                Phys. Let. B277 (1992) 1
\bibitem{bro}   V.Yu.Ponomarev et al., Phys. Rev. Lett. 72 (1994) 1168
\bibitem{AW79}  AA. Alder and K. Winther, Nucl. Phys. A319 (1979) 518
\bibitem{Ab}    M. Abramowitz and I.A. Stegun, Handbook of mathematical 
		functions (Dover publications, N.Y., 1965), p. 379 and 559 
\bibitem{SGII}  N. V. Giai, Suppl. Prog. Theor. Phys. 74-75 (1983) 330;
		N. V. Giai and H. Sagawa, Phys. Lett. B106 (1981) 379. 
\bibitem{Be89}  C. Benesh, B. Cook and J. Vary, Phys. Rev. C40 (1989) 1198.
\bibitem{ber96} C. A. Bertulani, V. Yu. Ponomarev and V. V. Voronov,
		preprint nucl-th/9606004 at preprint archive {\em
		http://xxx.lanl.gov/}
\bibitem{Gr}    I.S. Gradshteyn and I.M. Ryzhik, Tables of integrals, 
		series and products (Academic Press, N.Y., 1965).

\end{thebibliography}
\end{document}